\documentclass[9pt,twocolumn,twoside]{opticajnl}
\journal{opticajournal} % use for journal or Optica Open submissions

\setboolean{shortarticle}{true}
\usepackage{comment}
\usepackage{lineno}
\usepackage{subfigure}
\graphicspath{{figures/}}
\title{Quartic solitons of a mode-locked laser distributed model}

\author[1]{D. Malheiro}
\author[1,2,*]{M. Fac\~ao}
\author[3]{M. I. Carvalho}

\affil[1]{Departamento de F\'{i}sica, Universidade de Aveiro,
Campus Universit\'ario de Santiago, 3810-193 Aveiro, Portugal}
\affil[2]{I3N-Aveiro,Campus Universit\'ario de Santiago, 3810-193 Aveiro, Portugal}
\affil[3]{DEEC/FEUP and INESC TEC, Universidade do Porto, Rua Dr. Roberto Frias, 4200-465 Porto, Portugal}
\affil[*]{mfacao@ua.pt} %% email address is required

\begin{abstract}
Dissipative quartic solitons have gained interest in the field of mode-locked lasers for their energy-width scaling which allows the generation of ultrashort pulses with high energies. Pursuing the characterization of such pulses, here we found soliton solutions of a distributed model for mode-locked lasers in the presence of either positive or negative fourth order dispersion (4OD). We studied the impact the laser parameters may have on the profiles, range of existence and energy-width relation of the output pulses. The most energetic and narrowest solutions occur for negative 4OD, with the energy having an inverse cubic dependence with the width in most cases. Our simulations showed that the spectral filtering has the biggest contribution in the generation of short (widhts as low as 39 fs) and very energetic (392 nJ) optical pulses.
\end{abstract}

\setboolean{displaycopyright}{false} % Do not include copyright or licensing information in submission.

\begin{document}

\maketitle

Quartic solitons (QS) are shape-preserving pulses that occur in media with Kerr nonlinearity and fourth order dispersion (4OD), in the presence or in the absence of second order dispersion (group velocity dispersion - GVD). The conservative quartic solitons exist for negative 4OD and have been first studied in the 1990s \cite{karlsson94,akhmediev94,buryak95}
However, the interest on these solutions has increased more recently. Quartic solitons in silico-based slot waveguides were explored in \cite{roy2013}, soliton-like solutions were observed in photonic crystal waveguides with negative 4OD and negligible GVD (normal or anomalous) \cite{blanco-redondo16}. Tam \textit{et al.} \cite{tam19,tam20} have compiled the previous results of conservative models, including results in \cite{kruglov18},  by studying the nonlinear Schr\"odinger equation (NLS) plus negative 4OD, showing regions of existence of solitons with exponentially decaying tails, purely exponential or with oscillations. In 2017-19, quartic solitons were studied in dissipative Kerr optical cavities \cite{Bao2017, Taheri2019} and in 2020, a mode-locked laser using a cavity dispersion dominated by 4OD was demonstrated \cite{runge20}, thus opening the chapter of dissipative quartic solitons. Contrary to the conservative QS, the dissipative QSs have been shown to exist for both negative \cite{runge20,zhang22} and positive \cite{quian22} 4OD.

One of the advantages that has been pointed for the QS in ultrashort laser applications is its energy\textcolor{blue}{, $E$,} inversely proportional to the third power of temporal width\textcolor{blue}{, $\tau$}. This would yield very energetic short pulses. However, this energy-width law should be only universally valid for the conservative QSs. In fact, recent works on dissipative QSs have reported two different energy-width relations for the quartic soliton laser, Runge \textit{et al.} \cite{runge20} found \textcolor{blue}{$E\propto \tau^{-3}$} for negative 4OD  and Qian \textit{et al.} \cite{quian22} found \textcolor{blue}{$E\propto \tau^{3}$} for positive 4OD, both by varying the gain saturation energy. These contradictory results could be attributed to the sign of 4OD, but further studies are needed to establish if the law  \textcolor{blue}{$E\propto \tau^{-3}$} is always valid for dissipative QSs in the presence of negative 4OD. Indeed, it is well known that the characteristics of dissipative solitons are fixed by the equation parameters, an argument that was already referred for quartic solitons in mode-locked lasers in \cite{quian22}.

The works that have studied the QS laser, referred above, have used a lumped model, which has the advantage of a more accurate modelling of the real laser. The cubic quintic complex Ginzburg-Landau equation (CGLE) has also been extensively studied in connection with passively mode-locked lasers \cite{haus95,egorov09,zaviyalov10,grelu12}, with the advantage of having a lower number of parameters and allowing for semi-analytical approaches. Between those two approaches there is a distributed model, proposed in the works of Zaviyalov \textit{et al.}, which does not approximate the saturable absorber (SA) term that is present in the lumped model \cite{egorov09,zaviyalov10}. Here, we exploit the QS solutions of this distributed model including positive and negative 4OD, showing parameter ranges of existence, presenting their amplitude profiles, the dependence of the energy on temporal width and searching for effective ways to maximize pulse energy while simultaneously minimizing the width.

Let us consider an equation similar to equation (21) of \cite{zaviyalov10}, representing a distributed model for a mode-locked laser
\begin{multline}
i\frac{\partial W}{\partial z}-\frac{1}{2}\left(\beta_2+ig_0T_2^2\right)\frac{\partial^2W}{\partial t^2}+\frac{\beta_4}{24}\frac{\partial^4W}{\partial t^4}= \\
= \left(\frac{g_0-k_\text{OC}/L}{2}\right)W-\frac{i}{2}\frac{\delta_0 L_\text{SA}/L}{1+|W|^2/\bar{P}_\text{sat}}W-\bar{\gamma}|W|^2W, 
\label{eq:dist_model}
\end{multline}
where $t$ is the retarded time, $z$ is the propagation distance, $W(z,t)$ is the slowly varying pulse envelope, $\beta_2$ and $\beta_4$ are the second and fourth order dispersion parameters (the latter is introduced here), $g_0$ is the small signal gain, $T_2$ is the inverse linewidth of the parabolic gain and spectral filtering, $k_\text{OC}$ represents losses of the output coupler, $L$ is the cavity length and $\delta_0 L_\text{SA}$ is the modulation depth of the SA, with $\delta_0$ representing the small signal loss and $L_\text{SA}$ the length of the SA. The parameters $\bar{\gamma}$ and $\bar{P}_\text{sat}$ are effective parameters of the distributed model associated with the nonlinear parameter $\gamma$ and saturation power $P_\text{sat}$ of the SA, respectively, and given by $    \bar{\gamma}=\gamma(\exp(g_0L)-1)/g_0L$, $\bar{P}_\text{sat}=P_\text{sat}\exp(-g_0L)$. 
In this work we focus on the role played by $g_0$, $P_{sat}$, $T_2$ and $\beta_4$. All the following results were obtained with fixed $\delta_0 L_\text{SA} = 0.3$, $\gamma = 0.005$~W$^{-1}$m$^{-1}$, $L = 1$ m and $k_\text{OC} = -\text{ln}(0.3)$. 

We searched for stationary solutions of the form
$W(z, t) = F(t)e^{i\theta(t) + i\sigma z}$,
where both $F$ and $\theta$ are real and $\sigma$ is a propagation constant. Introducing this ansatz into \eqref{eq:dist_model} yields a system of nonlinear ordinary differential equations, which was solved using the Newton conjugate gradient method \cite{Yang2009, Yang2015} with hyperbolic secant inputs as well as soliton solutions from previous simulations. To verify that the solutions were stable, they were taken as inputs in \eqref{eq:dist_model} which was subsequently integrated through a pseudo-spectral method \cite{Fornberg1978} evolving in $z$ with a fourth order Runge-Kutta method, up to distances where the pulse was stationary. This approach is not very efficient in the identification of parameter regions where stable solutions exist and thus, we also calculated the eigenvalues of the linear stability operator for stationary solutions under \eqref{eq:dist_model}.

\begin{figure}[t] 
    \centering
    \subfigure[]{
    \centering\includegraphics[width = 0.5\linewidth]{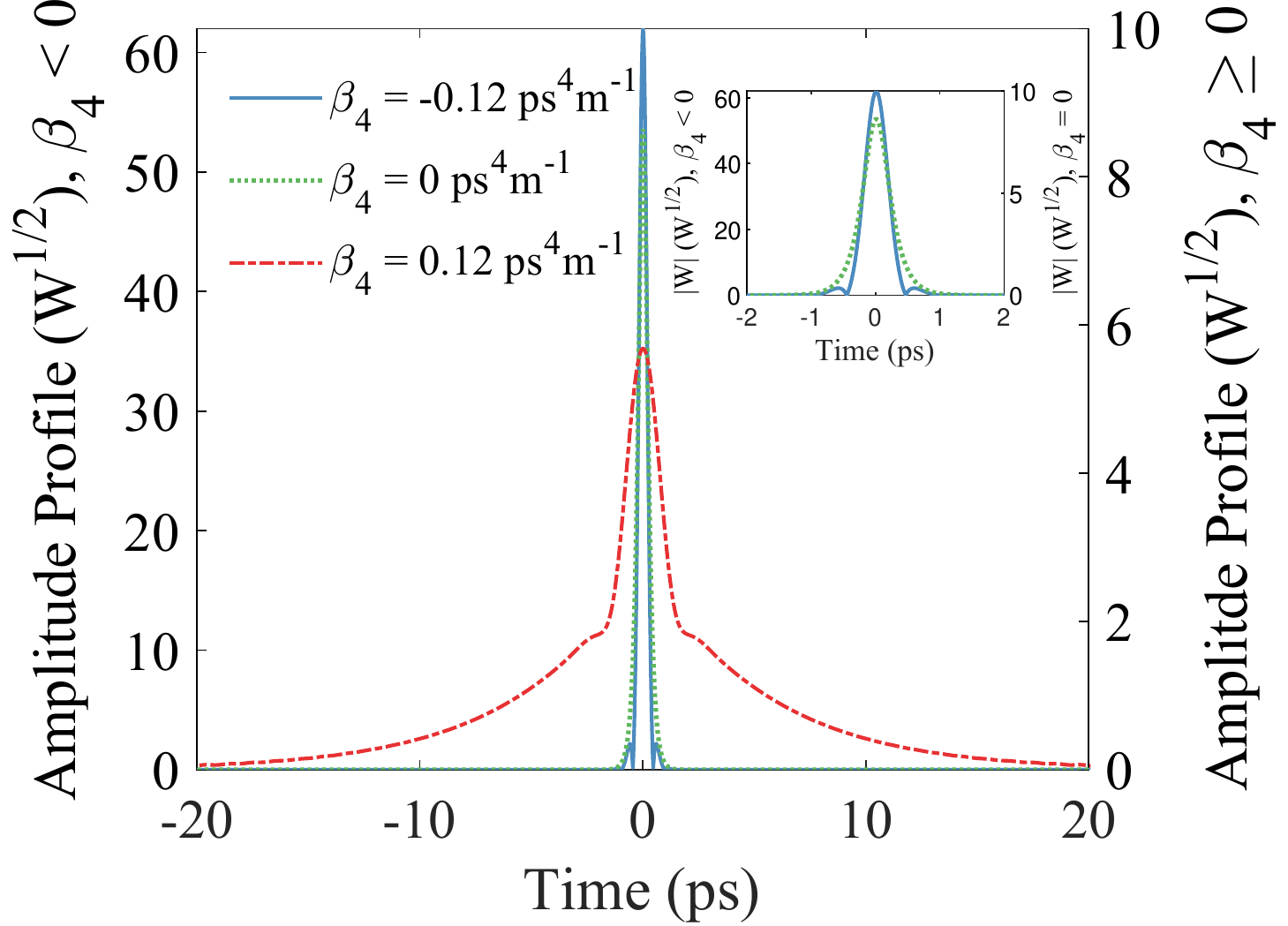}\label{fig:PROF_LIN_BETA4}}
    \subfigure[]{
    \centering\includegraphics[width = 0.46\linewidth]{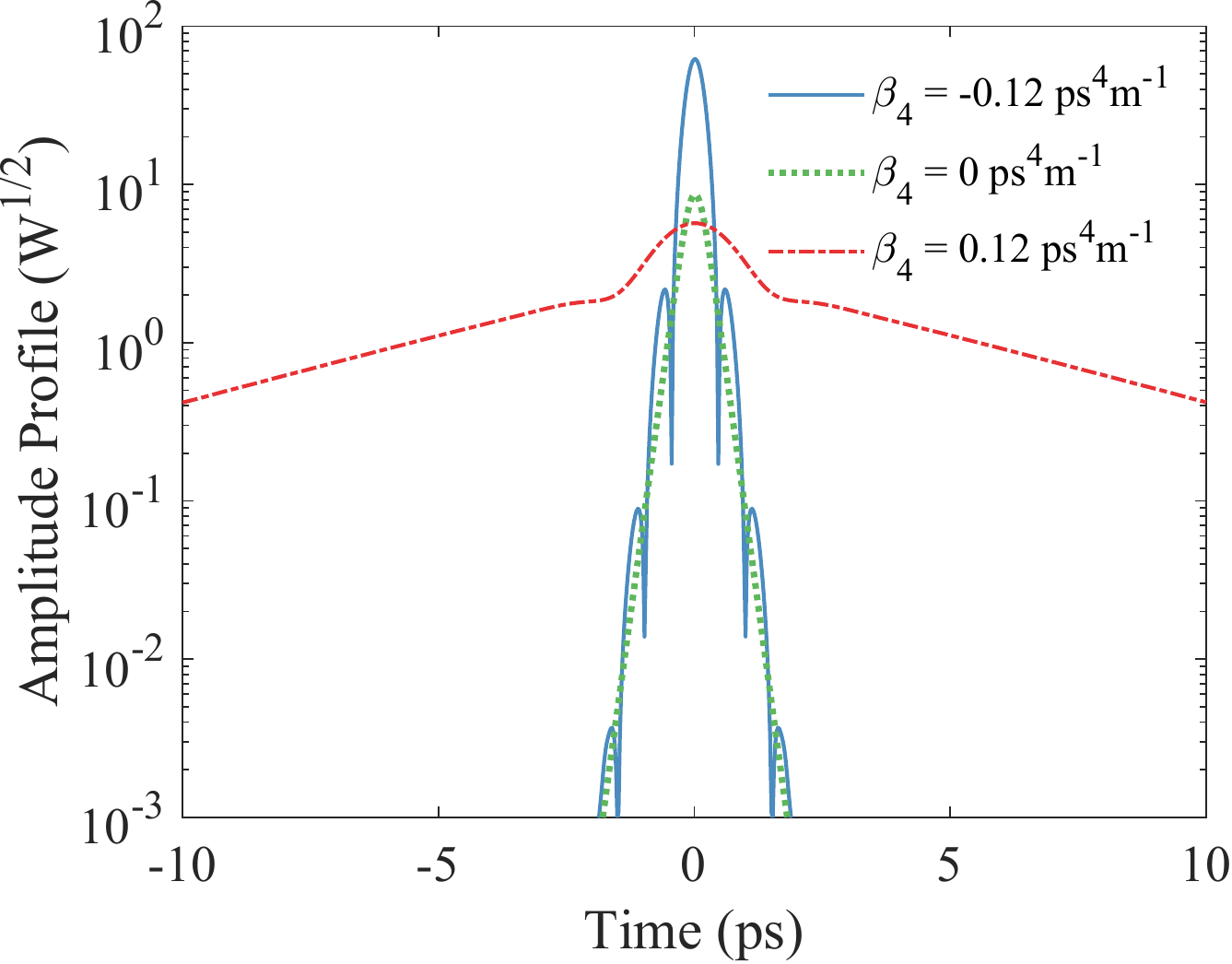}\label{fig:PROF_LOG_BETA4}}
    \subfigure[]{
    \centering\includegraphics[width = 0.49\linewidth]{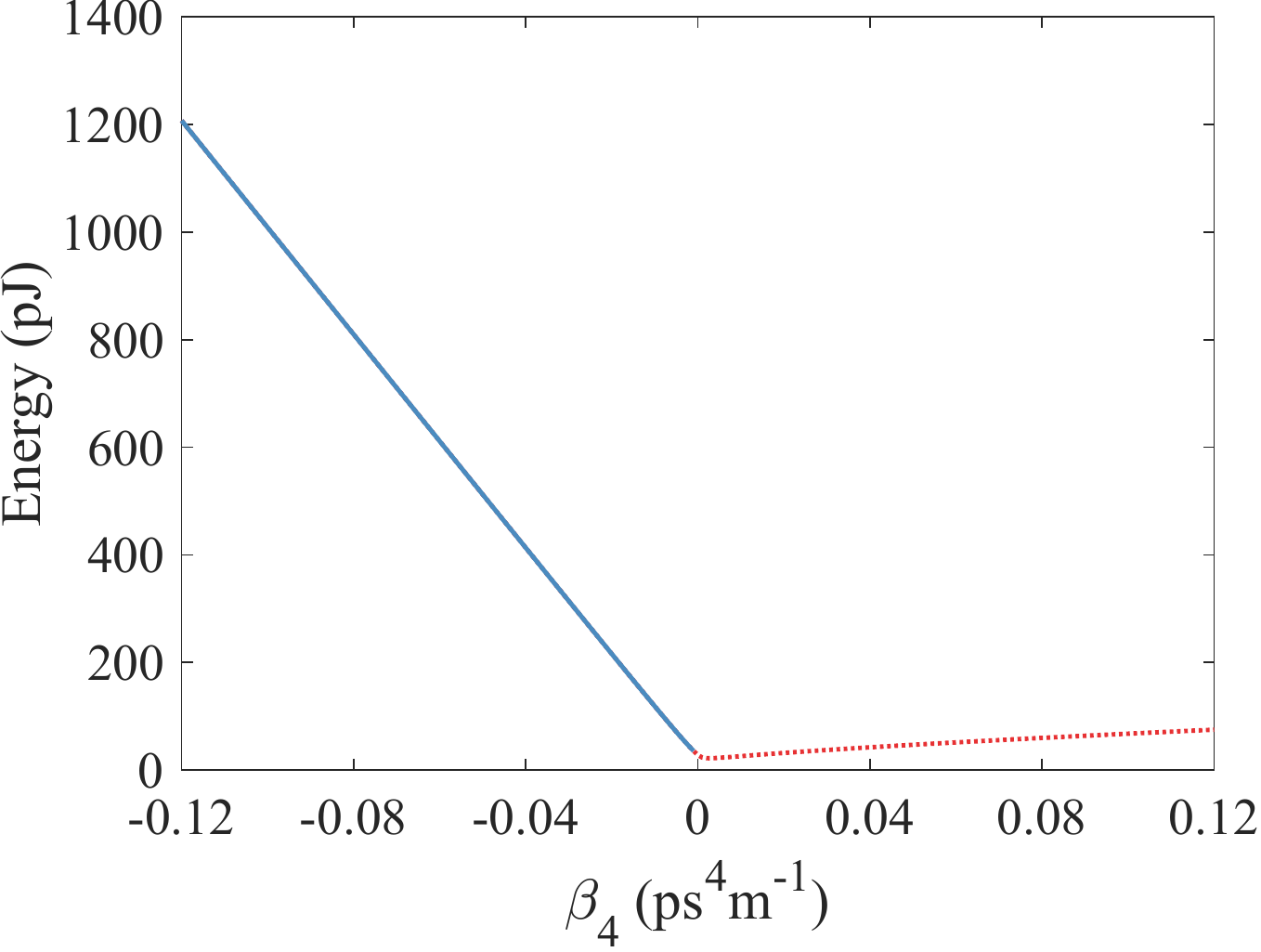}\label{fig:ENERGY_BETA4}}
    \subfigure[]{
    \centering\includegraphics[width = 0.47\linewidth]{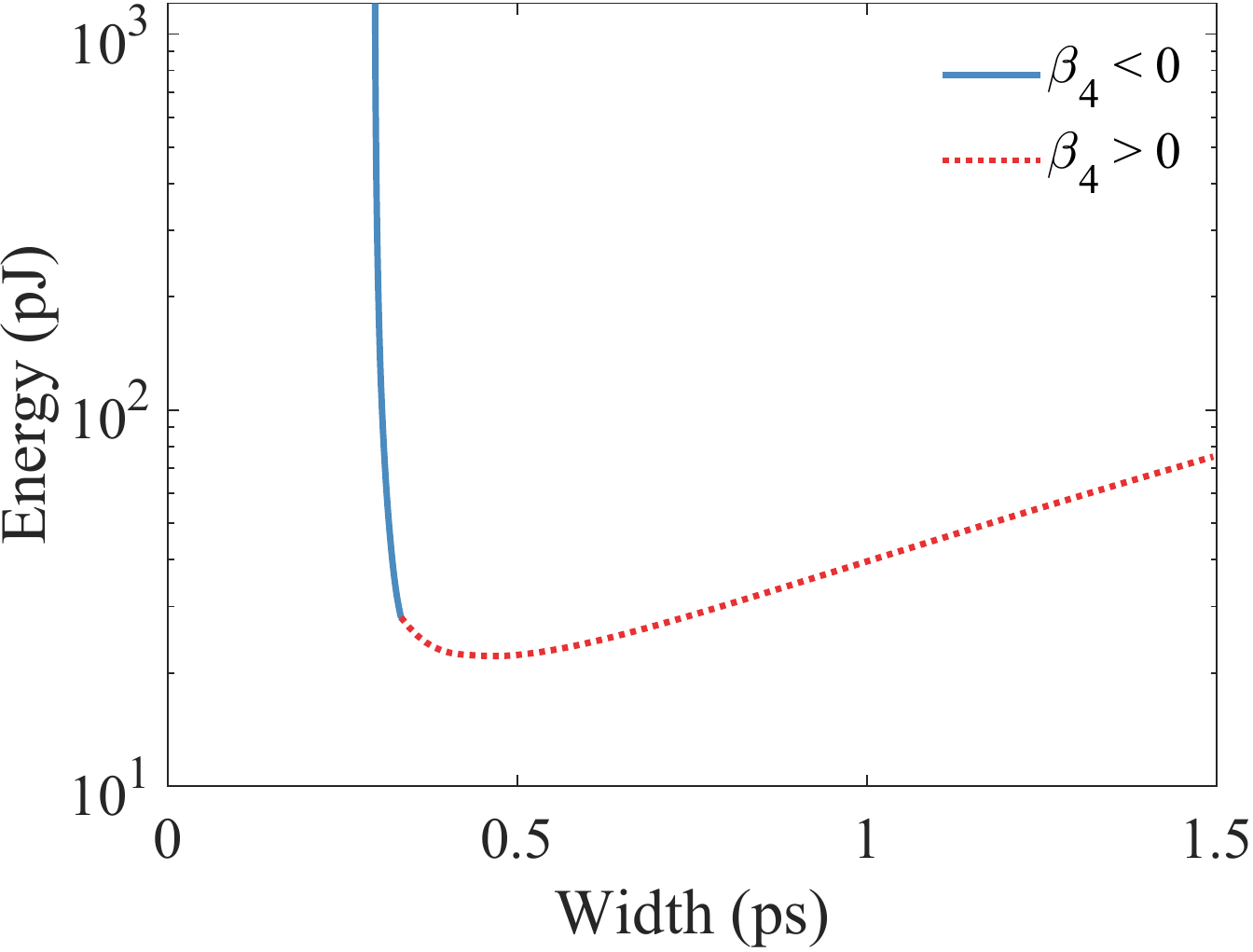}\label{fig:ENERGY_WIDTH_BETA4}}
    \caption{Pulse profiles in \subref{fig:PROF_LIN_BETA4} linear and \subref{fig:PROF_LOG_BETA4} logarithmic scale with negative, null and positive 4OD. The inset shows more detail in the pulse shape for $\beta_4 \leq 0$. Pulse energy \subref{fig:ENERGY_BETA4} and energy-width relation \subref{fig:ENERGY_WIDTH_BETA4} obtained from sweeping $\beta_4$ from $-0.12$ to $0.12~\text{ps}^4\text{m}^{-1}$. All solutions were obtained using $\beta_2  = -0.024~\text{ps}^2\text{m}^{-1}$, $g_0 = 1.45~\text{m}^{-1}$, $T_2 = 100~\text{fs}$ and $P_\text{sat} = 80~\text{W}$.}
\end{figure}

Stationary quartic soliton solutions were found for both positive and negative 4OD, with different signs of $\beta_4$ leading to different pulse shapes. As an example, in Fig. \ref{fig:PROF_LIN_BETA4} the pulse profiles for  $\beta_4 = \pm 0.012$ and $0~\text{ps}^4\text{m}^{-1}$ when $\beta_2 = -0.024~\text{ps}^2\text{m}^{-1}$, $g_0 = 1.45~\text{m}^{-1}$, $T_2 = 100~\text{fs}$ and $P_\text{sat} = 80~\text{W}$ are represented. The solution in the absence of 4OD takes the approximate shape of a hyperbolic secant, as is typical in GVD only solitons. For negative 4OD, the pulses are gaussian  with symmetrically oscillating tails, which are particularly noticeable in logarithmic scale (Fig. \ref{fig:PROF_LOG_BETA4}). For positive 4OD, the pulses come in the shape of a hyperbolic secant with broader tails. In logarithmic scale, these tails take the shape of straight lines, showing that they are, in fact, exponentially decaying. 

The pulse energy is also greatly influenced by the value of the 4OD coefficient.
Fig. \ref{fig:ENERGY_BETA4} shows that the energy increases with $\left|\beta_4\right|$, with the growth rate being greater when $\beta_4 < 0$. Fig. \ref{fig:ENERGY_WIDTH_BETA4} contains the energy-width relation of the pulses obtained by sweeping $\beta_4$ and keeping all other parameters constant. The pulses obtained with $\beta_4 < 0$ are, in general, the most energetic as well as the narrowest, with pulses as short as 300 fs having energies close to 1.2 nJ. For negative 4OD there is also a very sharp decrease of energy with an increase of the pulse width. In opposition, the energy increases with the width for positive $\beta_4$ but at a very slow rate. Therefore, the exploitation of negative 4OD in mode-locked laser cavities is much more favorable to achieve higher energies at lower pulse widths.

\begin{figure}[t]
    \centering
    \subfigure[]{
    \centering\includegraphics[width = 0.48\linewidth]{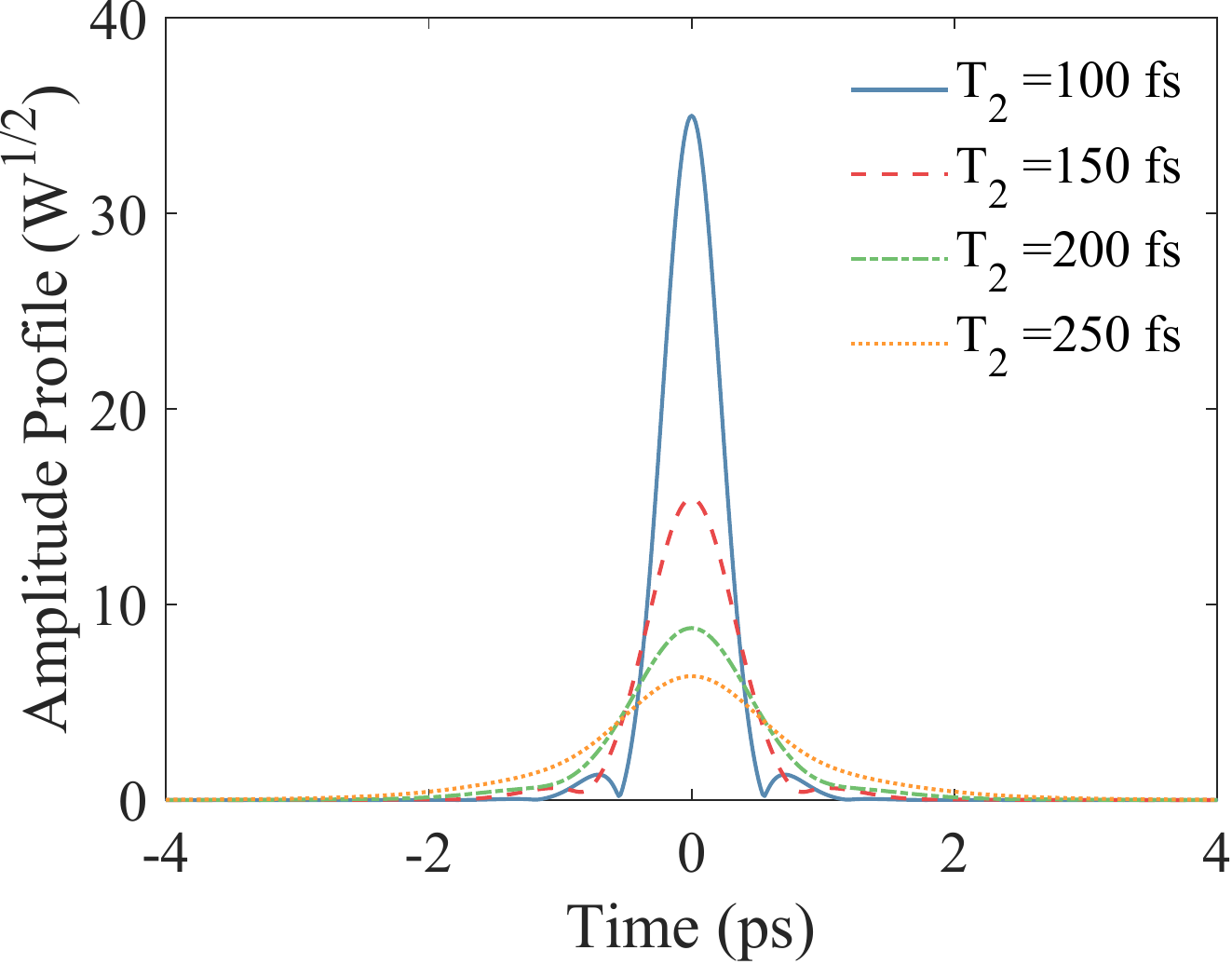}\label{fig:PROF_LIN_T2_NEG}}
    \centering
    \subfigure[]{
    \centering\includegraphics[width = 0.48\linewidth]{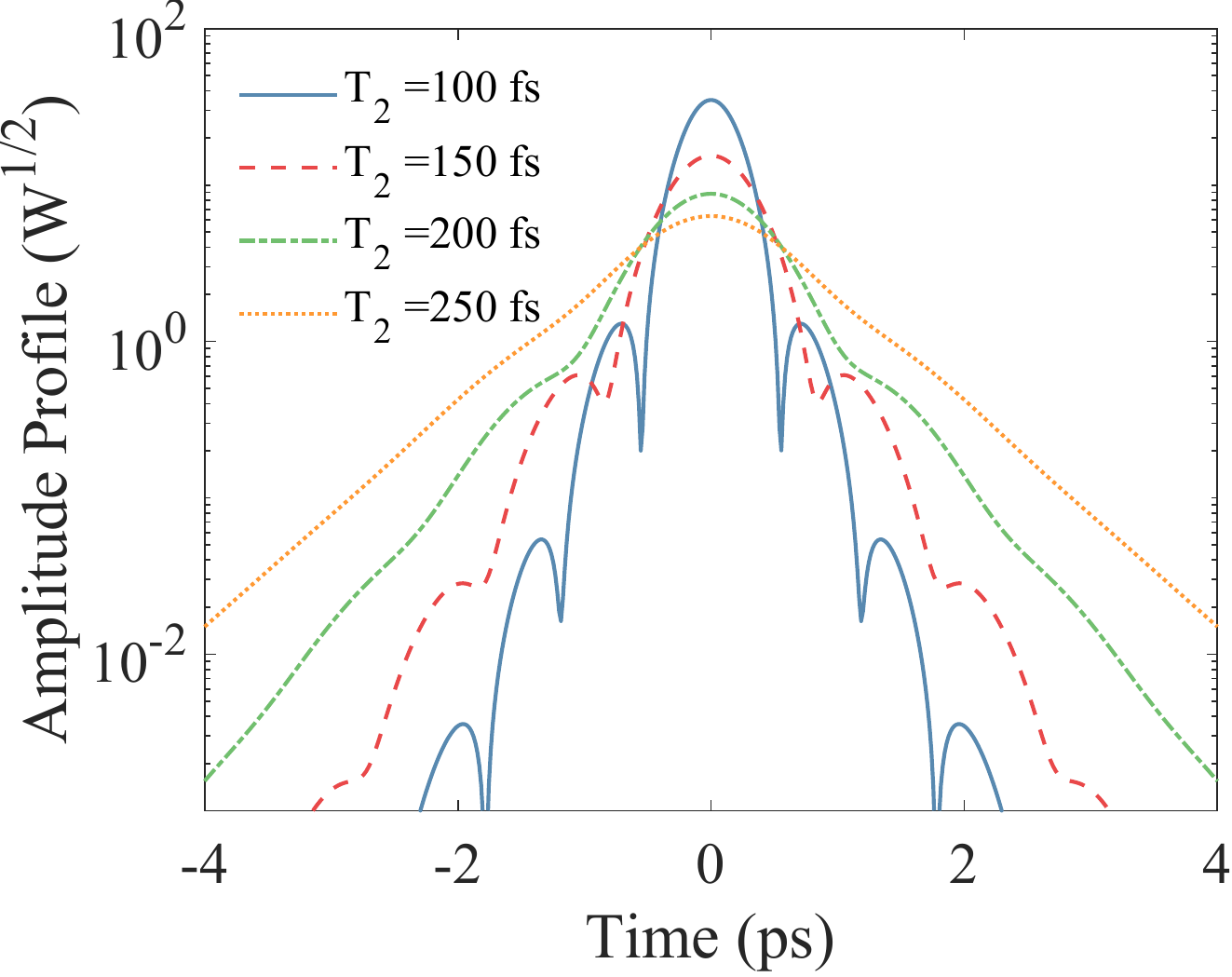}\label{fig:PROF_LOG_T2_NEG}}
   % \centering
   % \subfigure[]{
   % \centering\includegraphics[width = 0.48\linewidth]{figures/T2_SWEEP/PDE_PROF_LIN_POS_FOD.pdf}\label{fig:PROF_LIN_T2_POS_old}}
    %\centering
    %\subfigure[]{
    %\centering\includegraphics[width = 0.48\linewidth]{figures/T2_SWEEP/PDE_PROF_LOG_POS_FOD.pdf}\label{fig:PROF_LOG_T2_POS_old}}
       \centering
    \subfigure[]{
    \centering\includegraphics[width = 0.48\linewidth]{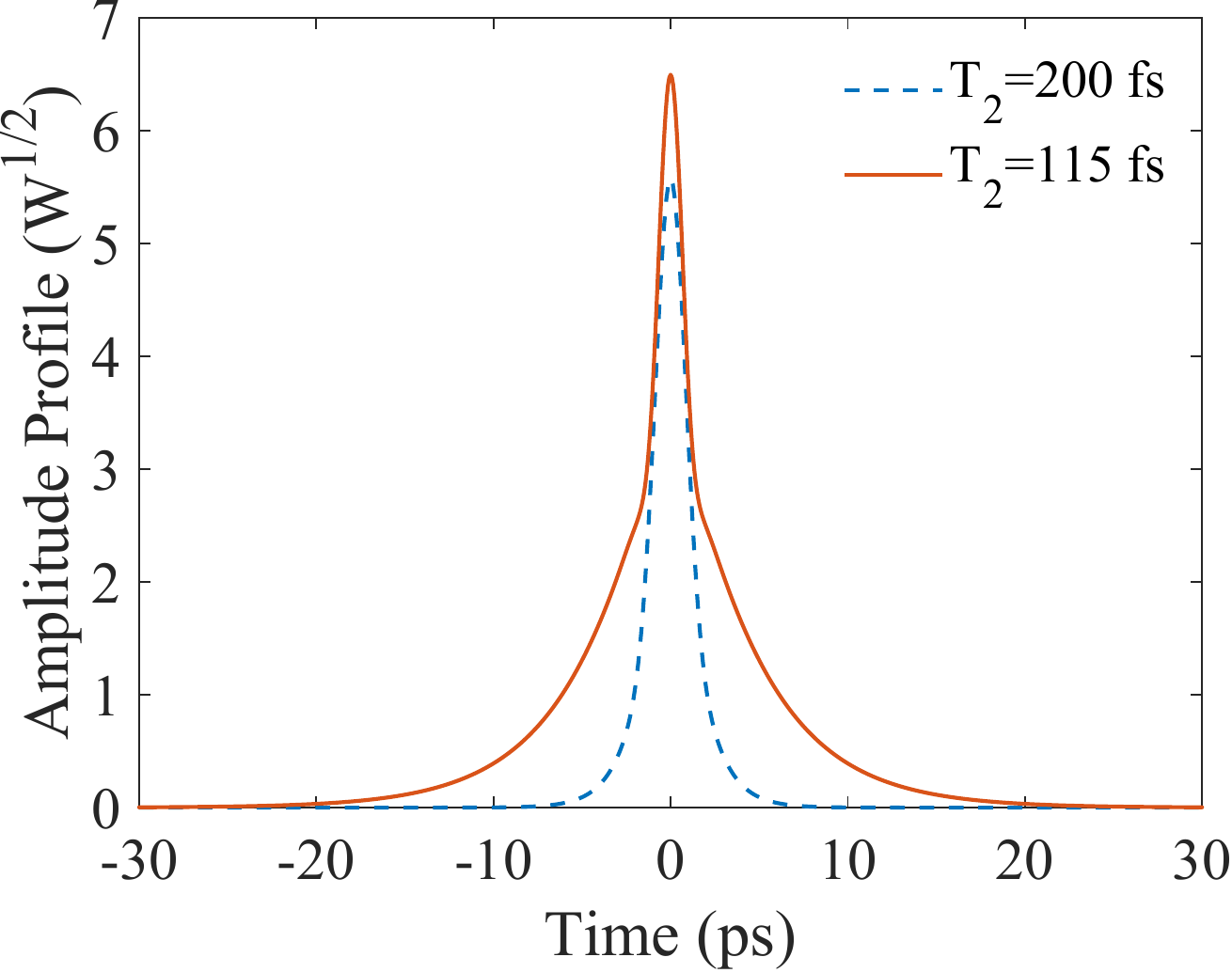}\label{fig:PROF_LIN_T2_POS}}
    \centering
    \subfigure[]{
    \centering\includegraphics[width = 0.48\linewidth]{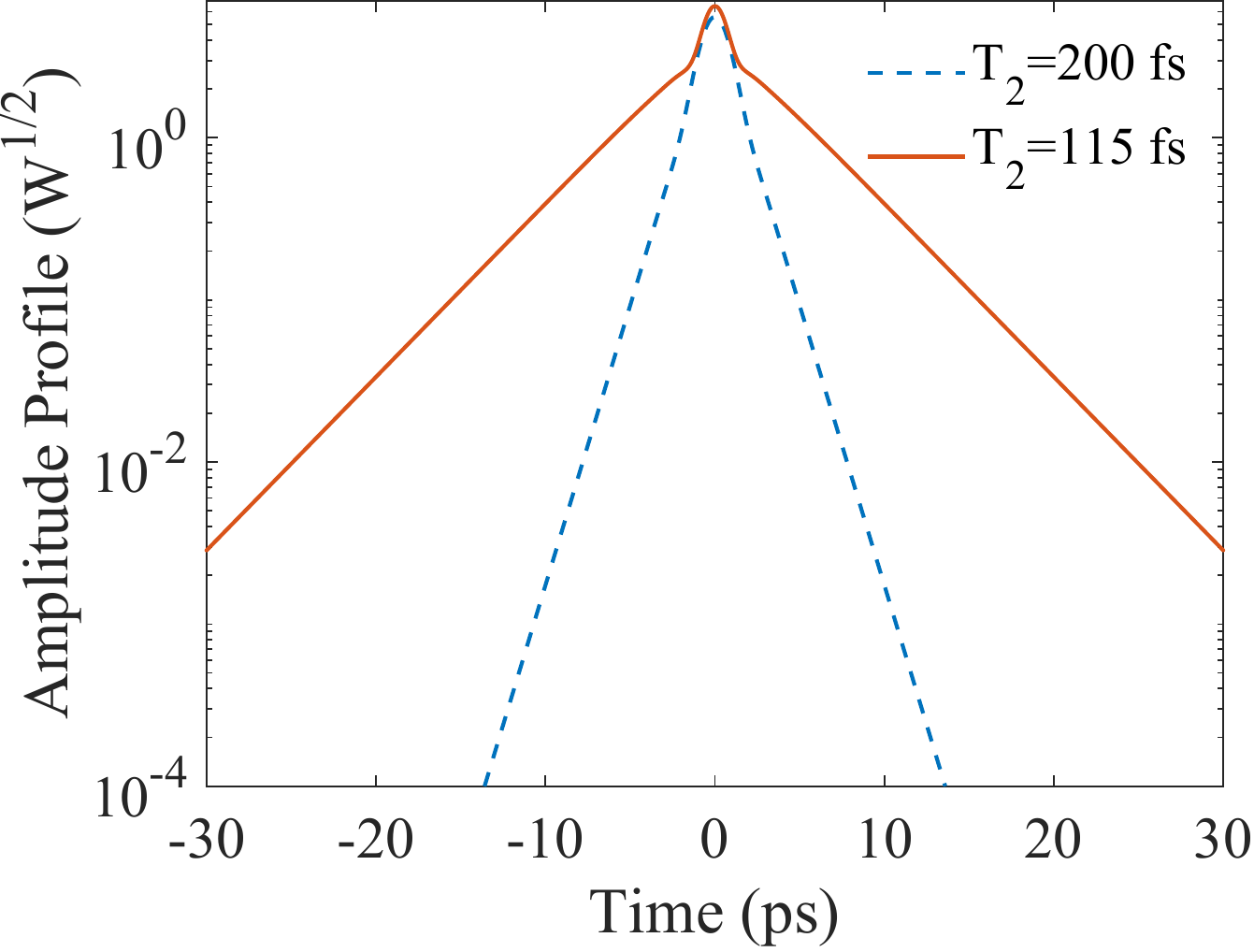}\label{fig:PROF_LOG_T2_POS}}
    \caption{Pulse shapes for different $T_2$ for \subref{fig:PROF_LIN_T2_NEG}, \subref{fig:PROF_LOG_T2_NEG} $\beta_4=-0.08~\text{ps}^4\text{m}^{-1}$, $P_\text{sat}=3~\text{W}$ and $g_0 = 1.36~\text{m}^{-1}$,  and \subref{fig:PROF_LIN_T2_POS}, \subref{fig:PROF_LOG_T2_POS} $\beta_4 = 0.08~\text{ps}^4\text{m}^{-1}$, $P_\text{sat} = 80~\text{W}$ and $g_0 = 1.48~\text{m}^{-1}$, 
    in linear and logarithmic scale, respectively. In both cases, $\beta_2 = 0$.} 
\end{figure}

The sign of 4OD is not the only factor defining the shape of the pulses. All of the parameters under investigation here were found to have some effect on the pulse shape, amplitude and width. Fig. \ref{fig:PROF_LIN_T2_NEG} shows the impact of growing $T_2$ in the shape of pulses obtained with $\beta_4 = -0.08~\text{ps}^4\text{m}^{-1}$. For $T_2 = 100\text{ fs}$ (blue solid curve), the pulse takes the same shape as the one with negative 4OD in Fig. \ref{fig:PROF_LIN_BETA4}. As $T_2$ increases to 150 fs, and further so to 200 fs (red dashed and green dash-dot curves respectively), the oscillations in the tails begin to flatten. For $T_2 = 250~\text{fs}$ the pulse becomes hyperbolic secant shaped, maintaining this shape for all greater $T_2$ values where stationary solutions exist.  The flattening of the tail oscillations is further evidenced in logarithmic scale (Fig. \ref{fig:PROF_LOG_T2_NEG}), with the oscillations in the tails becoming less and less noticeable until the tails become approximately straight lines. A similar behaviour exists in the presence of positive 4OD. In Fig. \ref{fig:PROF_LIN_T2_POS}, considering $\beta_4 = 0.08~\text{ps}^4\text{m}^{-1}$, $g_0 = 1.48~\text{m}^{-1}$, $P_\text{sat} = 80~\text{W}$ and $T_2 = 115~\text{fs}$, a pulse with exponentially decaying tails is obtained, similar to the one in Fig. \ref{fig:PROF_LIN_BETA4}. Changing $T_2$ to 200 fs results in a hyperbolic secant shaped pulse.  Fig. \ref{fig:PROF_LOG_T2_POS} shows the different pulse shapes in logarithmic scale, highlighting a much more rapid decay in the tails for $T_2 = 200~\text{fs}$. 

\begin{figure}[t]%Energias com os Parâmetros
    \centering
    \subfigure[]{
    \centering\includegraphics[width = 0.48\linewidth]{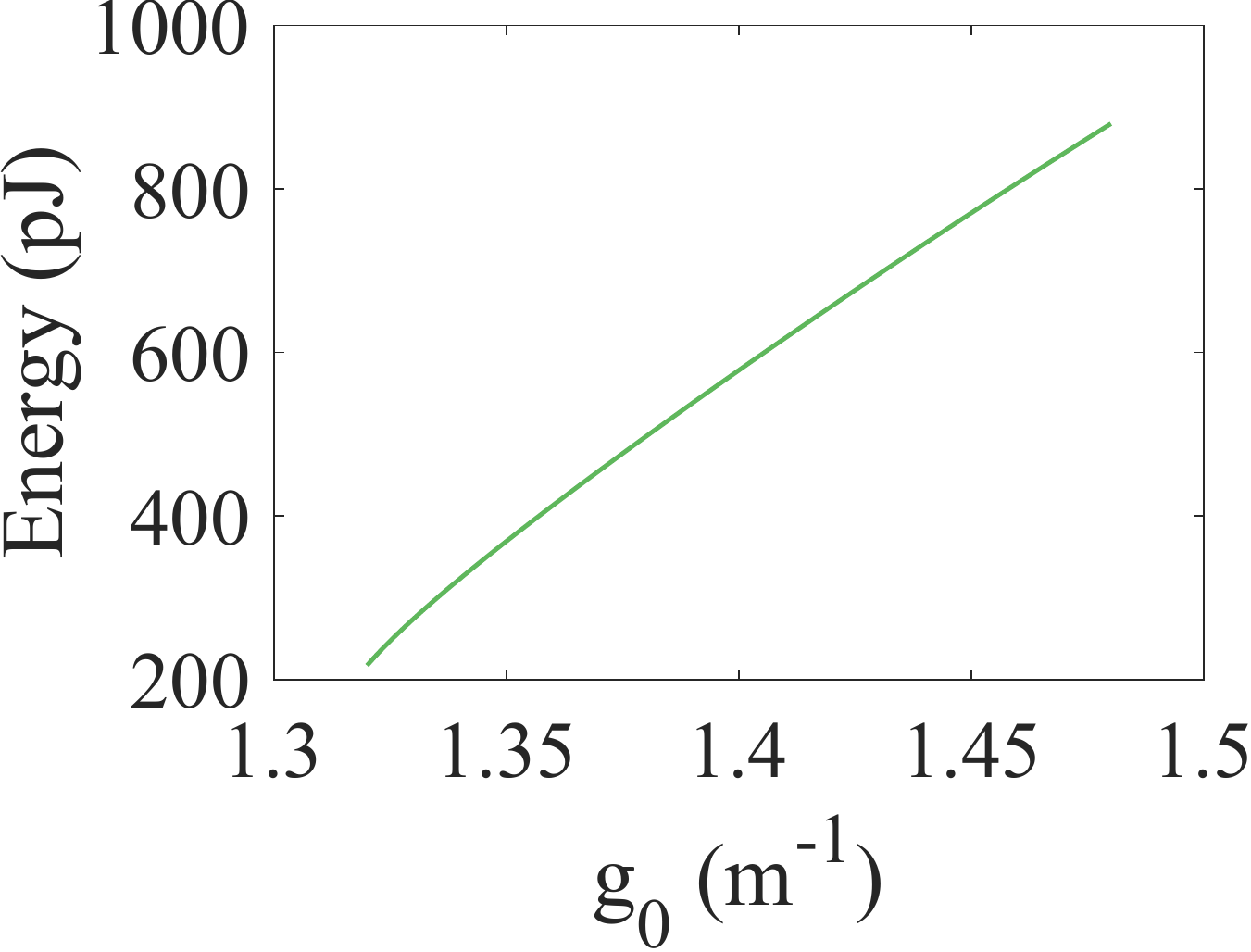}\label{fig:E_G0_NEG_FOD}}
     \subfigure[]{
    \centering\includegraphics[width = 0.48\linewidth]{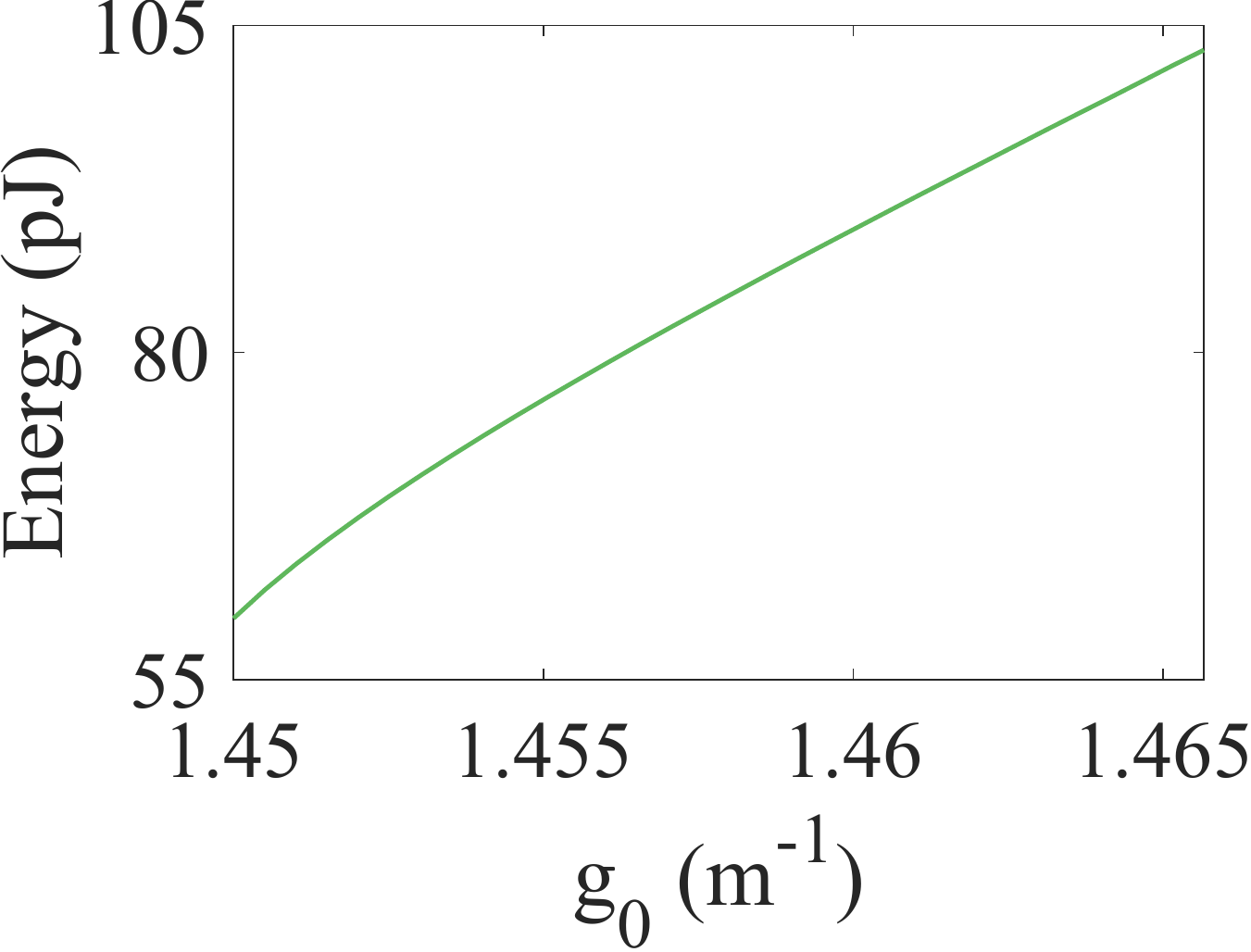}\label{fig:E_G0_POS_FOD}}
    \subfigure[]{
    \centering\includegraphics[width = 0.48\linewidth]{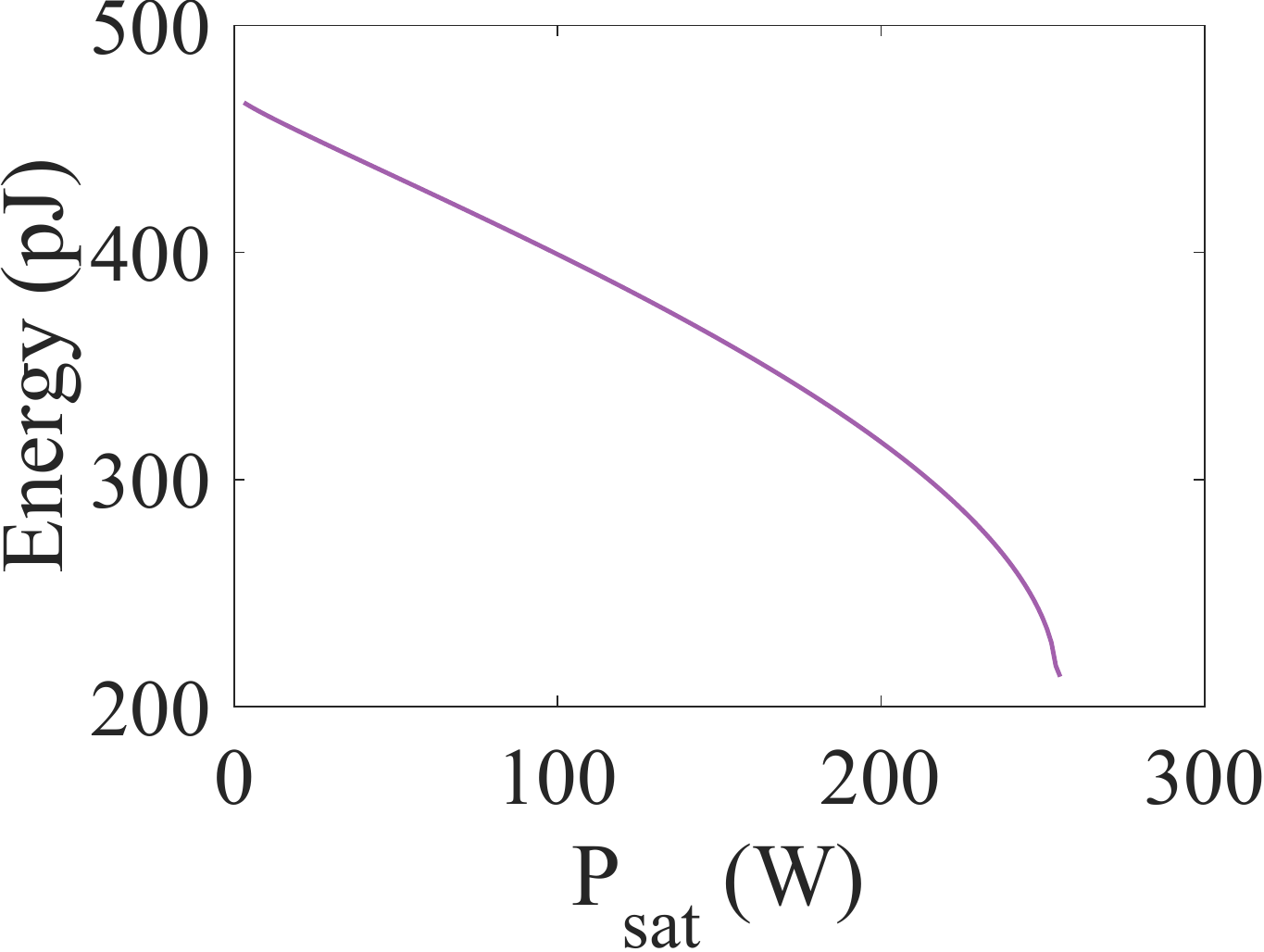}\label{fig:E_PSAT_NEG_FOD}}
     \subfigure[]{
    \centering\includegraphics[width = 0.48\linewidth]{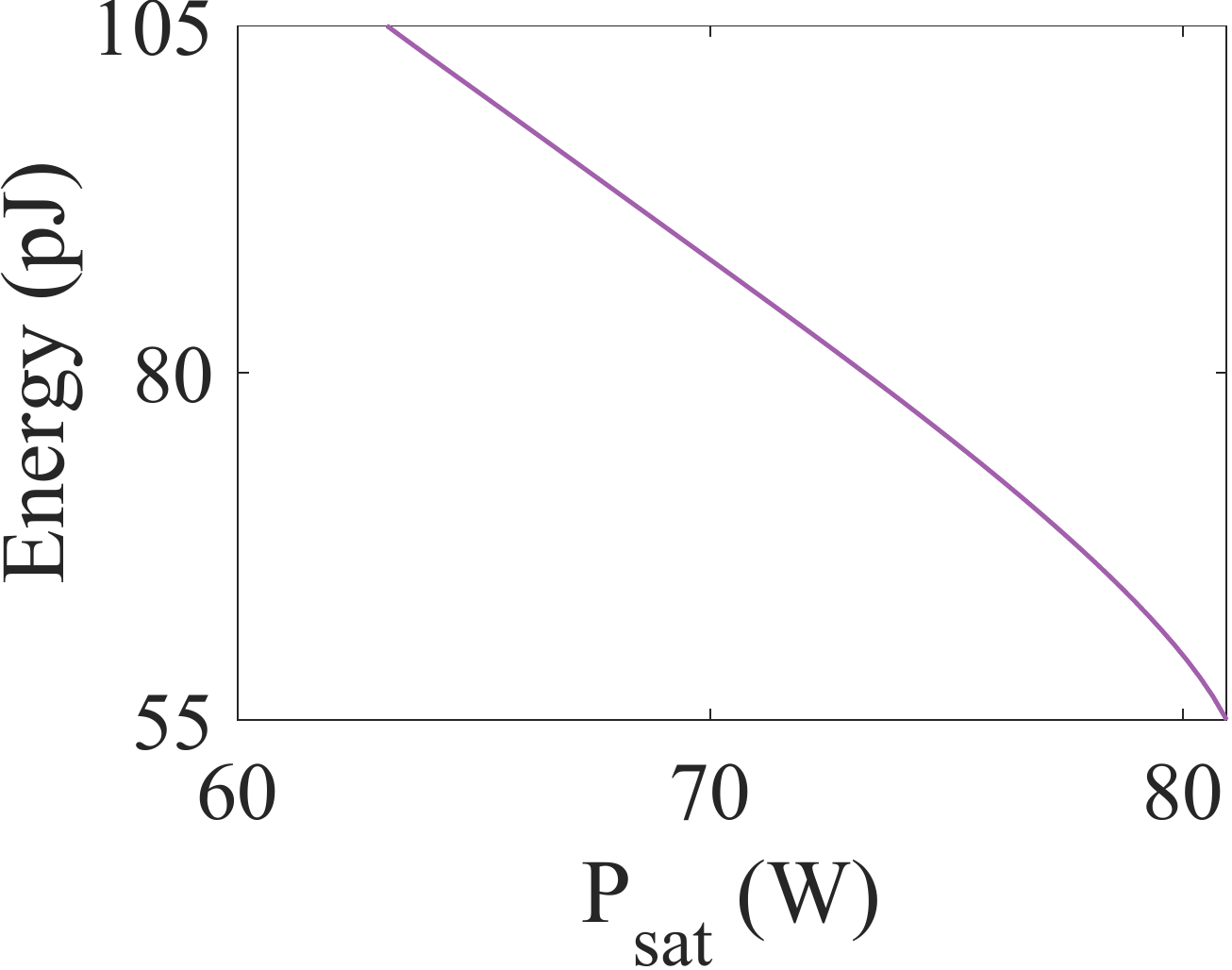}\label{fig:E_PSAT_POS_FOD}}
    \subfigure[]{
    \centering\includegraphics[width = 0.48\linewidth]{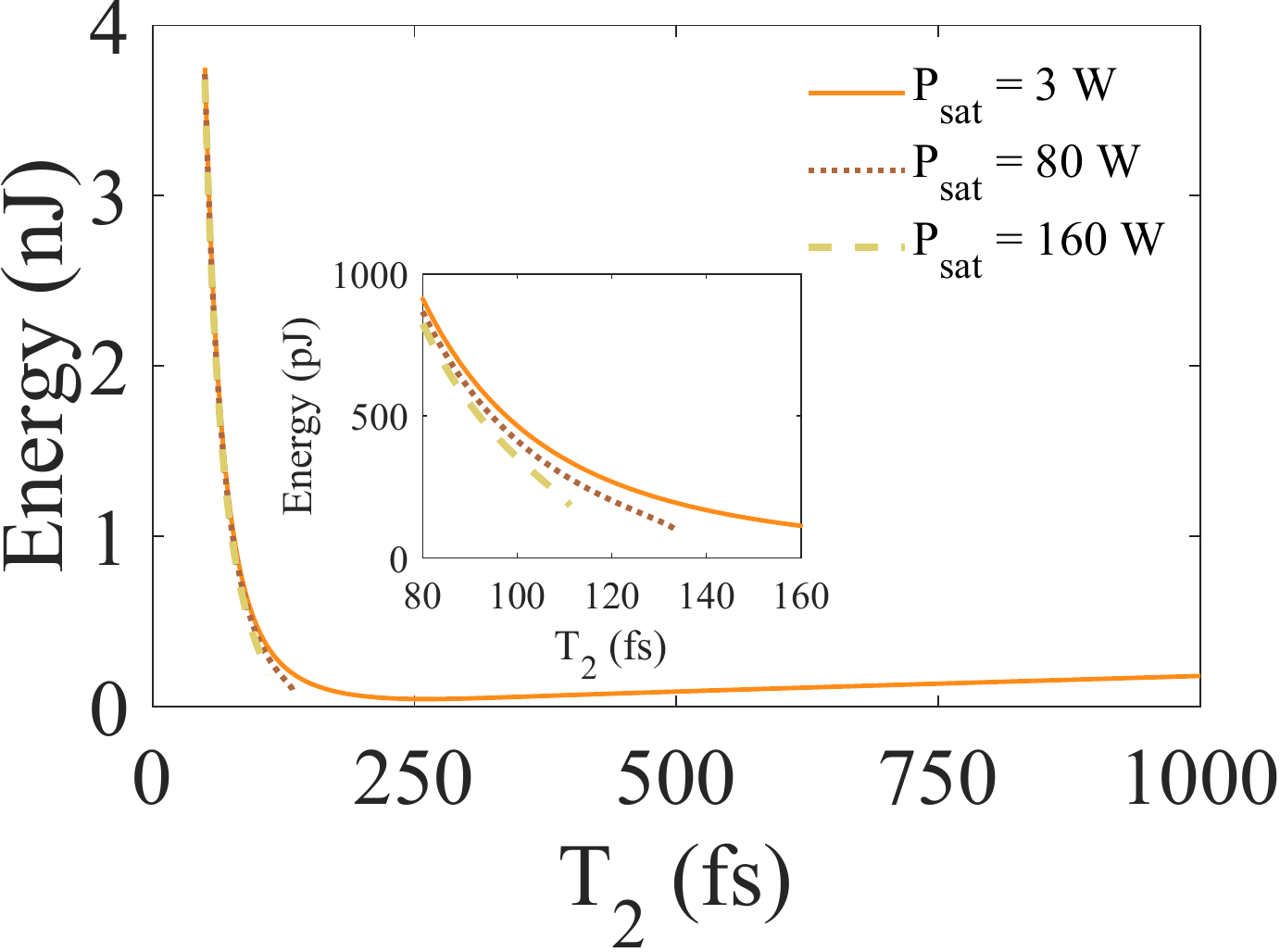}\label{fig:E_T2_NEG_FOD}}
    \subfigure[]{
    \centering\includegraphics[width = 0.48\linewidth]{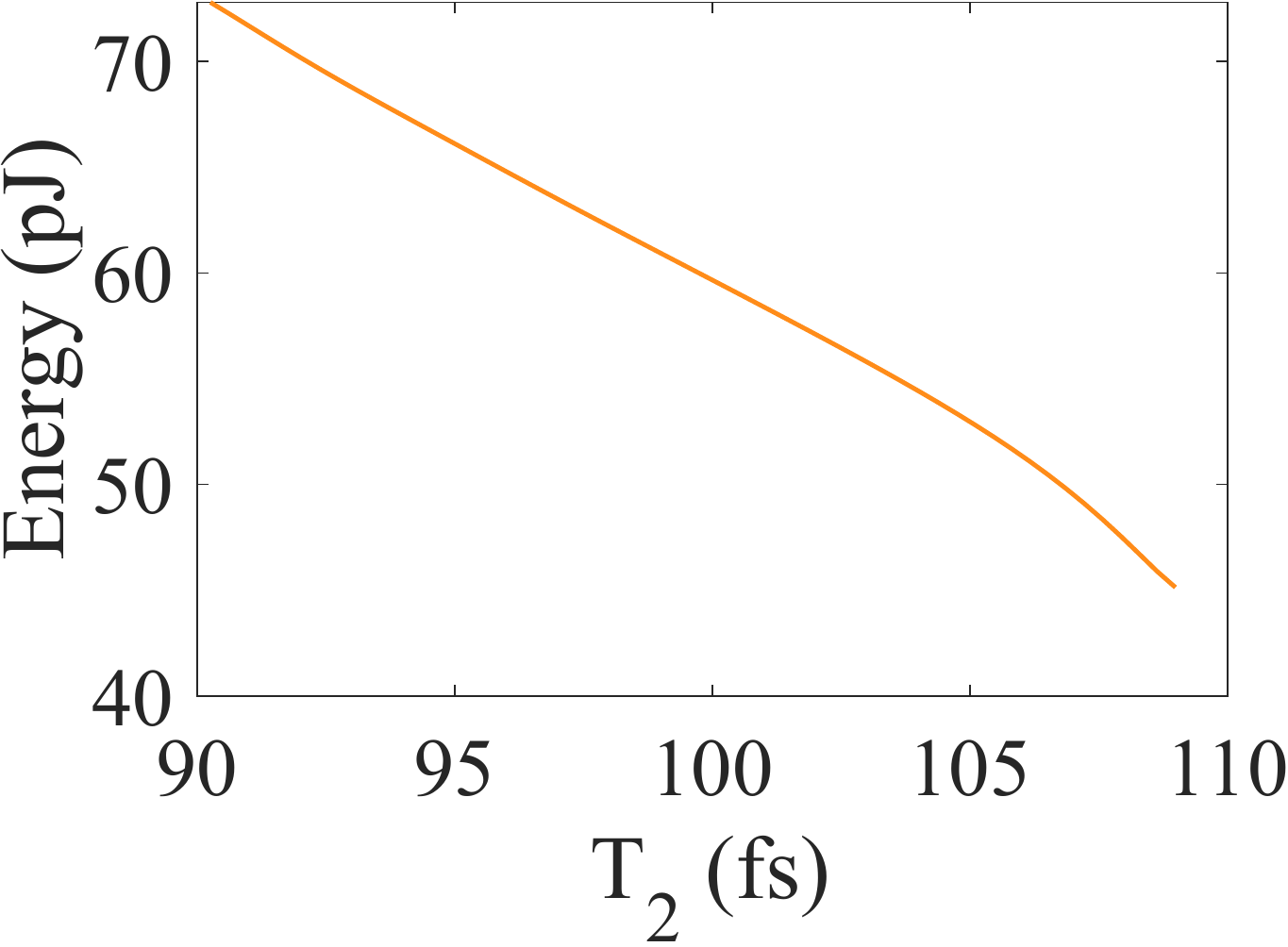}\label{fig:E_T2_POS_FOD}}
    \caption{Energy dependence with $g_0$ (top row), $P_\text{sat}$ (middle row) and $T_2$ (bottom row) for $\beta_4 = -0.08~\text{ps}^4\text{m}^{-1}$ (left) and $\beta_4 = 0.08~\text{ps}^4\text{m}^{-1}$ (right).  When constant, $P_\text{sat} = 80~\text{W}$ and $T_2 = 100~\text{fs}$ were considered in all cases, while $g_0$ took values of $1.36~\text{m}^{-1}$ for $\beta_4 < 0$ and of $1.45~\text{m}^{-1}$ for $\beta_4 > 0$.}
    \label{fig:energy_parameter}
\end{figure}

To design a mode-locked fiber laser capable of emitting high energy pulses, it is important to find how the different laser parameters impact the pulse energy. In Fig. \ref{fig:energy_parameter}, the energy dependence with $g_0$, $P_\text{sat}$ and $T_2$ is plotted for $\beta_4 < 0$ (Figs. \ref{fig:E_G0_NEG_FOD}, \ref{fig:E_PSAT_NEG_FOD}, \ref{fig:E_T2_NEG_FOD}, respectively) and $\beta_4 > 0$ (Figs. \ref{fig:E_G0_POS_FOD}, \ref{fig:E_PSAT_POS_FOD}, \ref{fig:E_T2_POS_FOD}, respectively). It immediately becomes apparent, reinforcing the results from Fig. \ref{fig:ENERGY_BETA4}, that the pulses with highest energies occur in the negative 4OD regime. In fact, the highest energy obtained in a $g_0$ sweep for negative 4OD is about 9 times greater than the one found for positive 4OD. More striking is the case of the $T_2$ sweep, where the highest energy with $\beta_4 < 0$ is approximately 3 orders of magnitude larger than the one for $\beta_4 > 0$. It is also clear that the highest energies occur for higher $g_0$ values as well as low $P_\text{sat}$ and $T_2$ values. It is important to note however, that the choice of just one of these parameters is enough to completely alter the range where the other parameters allow stationary solutions. As an example, in Fig. \ref{fig:E_T2_NEG_FOD}, with a saturation power of 3 W, stationary solutions were found from $T_2 = 50$ fs to $T_2 > 1$ ps. However, by simply increasing $P_\text{sat}$ to 80 W, the maximum $T_2$ value reached was of approximately 140 fs, and became of $\sim 110$ fs when $P_\text{sat} = 160~\text{W}$ was considered (see inset). The use of positive 4OD also leads to a much shorter valid $g_0$, $P_\text{sat}$ and $T_2$ ranges for the existence of stationary soliton solutions of \eqref{eq:dist_model}. Finally, Fig. \ref{fig:energy_parameter} also shows that the different parameters also have very different impacts in the output pulse energy. For instance, with $\beta_4 < 0$, very small changes for $T_2 \lesssim 150$ fs, lead to drastic variations on the pulse energy, while small increases in $P_\text{sat}$ do not necessarily lead to very significant energy decreases. In fact, in Fig. \ref{fig:E_T2_NEG_FOD}, the change in $P_\text{sat}$ altered the soliton stability range, but changes in energy were not very significant.

\begin{figure} [t]%Energia vs Largura
    \centering
    \subfigure[]{
    \centering\includegraphics[width = 0.486\linewidth]{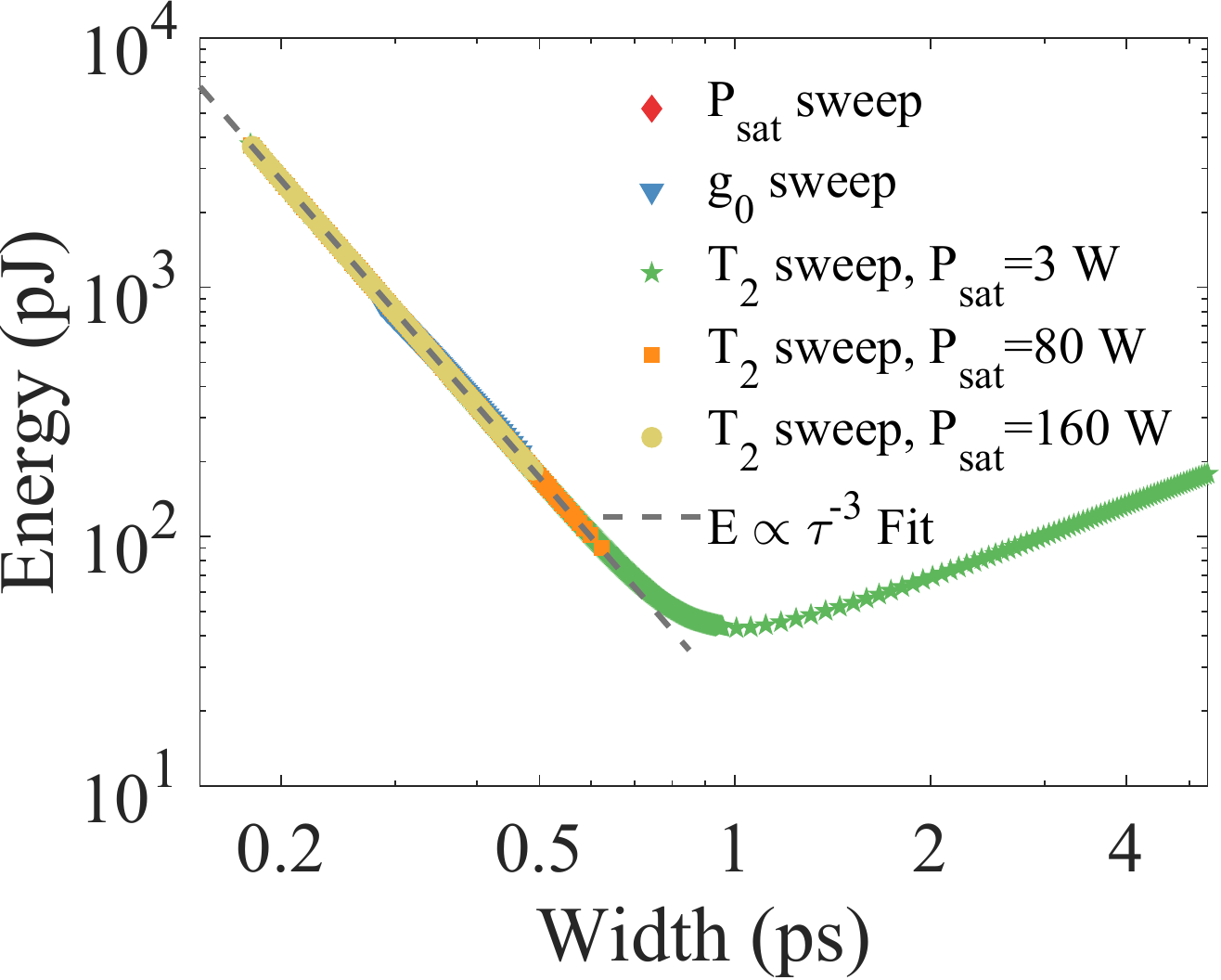}\label{fig:EW_NEG_FOD}}
    \subfigure[]{
    \centering\includegraphics[width = 0.486\linewidth]{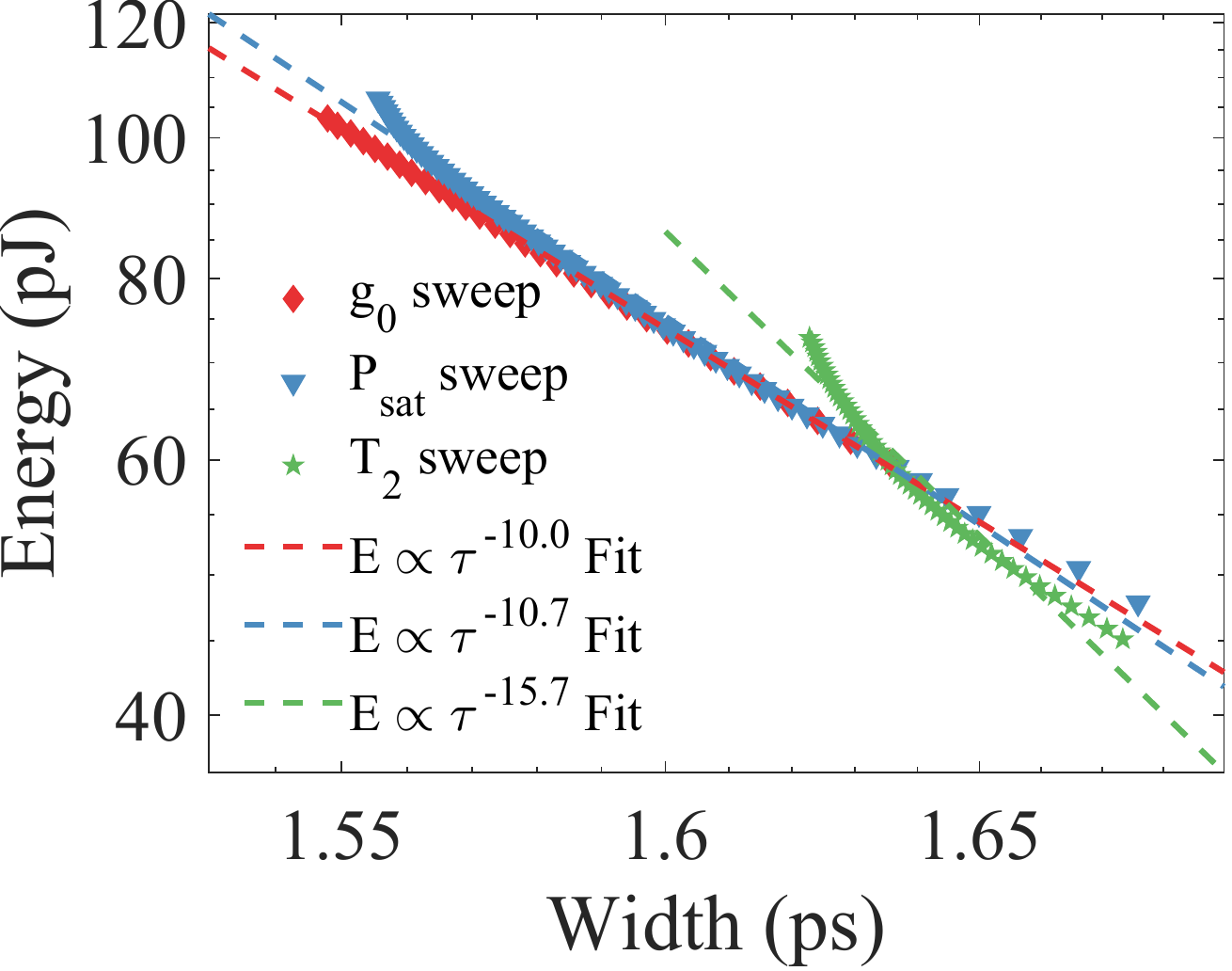}\label{fig:EW_POS_FOD}}
    \caption{Energy-width curves using logarithm base 10 scale and associated fittings, obtained by sweeping $g_0$, $P_\text{sat}$ and $T_2$ for $\beta_4 = -0.08~\text{ps}^4\text{m}^{-1}$ \subref{fig:EW_NEG_FOD} and $\beta_4 = 0.08~\text{ps}^4\text{m}^{-1}$ \subref{fig:EW_POS_FOD}. When constant, values of $T_2 = 100~\text{fs}$ were considered , while $g_0$ took values of $1.36~\text{m}^{-1}$ for $\beta_4 < 0$ and of $1.45~\text{m}^{-1}$ for $\beta_4 > 0$ and $P_\text{sat}$ is 80 W in all cases not referred in the legends.}
    \label{fig:ENERGY-WIDTH}
\end{figure}

As noted above, quartic soliton lasers have gained interest due to their ability to generate highly energetic pulses with short widths. In the distributed-model, it was found, in most cases, that the energy does scale inversely with the width cubed, in agreement with previous studies on the subject using other models \cite{tam19, runge20}. In Fig. \ref{fig:ENERGY-WIDTH} the energy-width relation of the soliton solutions of \eqref{eq:dist_model} are presented, with Fig. \ref{fig:EW_NEG_FOD} containing the results for the sweeps of $g_0$, $P_\text{sat}$ and $T_2$ with negative 4OD respectively, while in Fig. \ref{fig:EW_POS_FOD} the same curves are plotted for positive 4OD. Interestingly, most of the energy-width points for negative 4OD fall approximately in the same curve that, in fact, follows the relation $E\propto \tau^{-3}$ (see gray-dashed line). The exception is for one of the branches occurring for $P_\text{sat}=3$ W while $T_2$ is swept above 150~fs, for which the energy increases with width. For positive 4OD however, this relation is not verified, nor is the cubic dependence predicted in \cite{quian22}. In fact, our curve fittings to data for $\beta_4 > 0$ yielded different relations as may be observed by the dashed lines of Fig. \ref{fig:EW_POS_FOD} that were obtained by fitting the data to the function $a\tau^b$ (where $a$ and $b$ are fitting coeficients). The pulses for $\beta_4 > 0$ not only have lower energies but they are also much wider. In this case, the shortest and most energetic pulses have widths close to 1.6 ps and energies close to 100 pJ, while the pulses with $\beta_4 < 0$ reach energies in the nJ range, and widths close to 300 fs.

The results presented in Fig. \ref{fig:energy_parameter} indicate how a single parameter, $g_0$, $P_{sat}$ or $T_2$, could be manipulated to obtain pulses with higher energy. Nevertheless, the maximization of the pulse energy must take into account the simultaneous variation of all these parameters. Not only is the dependence of the pulse energy on these parameters quite different but, more importantly, stable solutions only exist within a certain parameter region.  
In effect, an increase of $P_\text{sat}$ implies that the SA saturates for higher power and thus, the nonlinear gain decreases. For a higher value of $P_\text{sat}$, the nonlinear gain may not be sufficient to counter balance other losses and thus, soliton solutions are not found. This same effect also occurs with lower $g_0$ values, which lead to higher effective saturation powers, $\bar{P}_\text{sat}$. For higher $g_0$ values, the linear losses ($-g_0/2 + k_\text{OC}/2L + \delta_0 L_\text{SA}/2L$) tend to zero and linear gain would produce unstable background. On the other hand, lower values of $P_\text{sat}$ lead to a sharp increase in the nonlinear gain which would be unbalanced by the other parameters, also leading to unstable solitons.

\begin{figure} %Estabilidade e Casos Extremos
 \centering
    \subfigure[]{
    \centering\includegraphics[width = 0.48\linewidth]{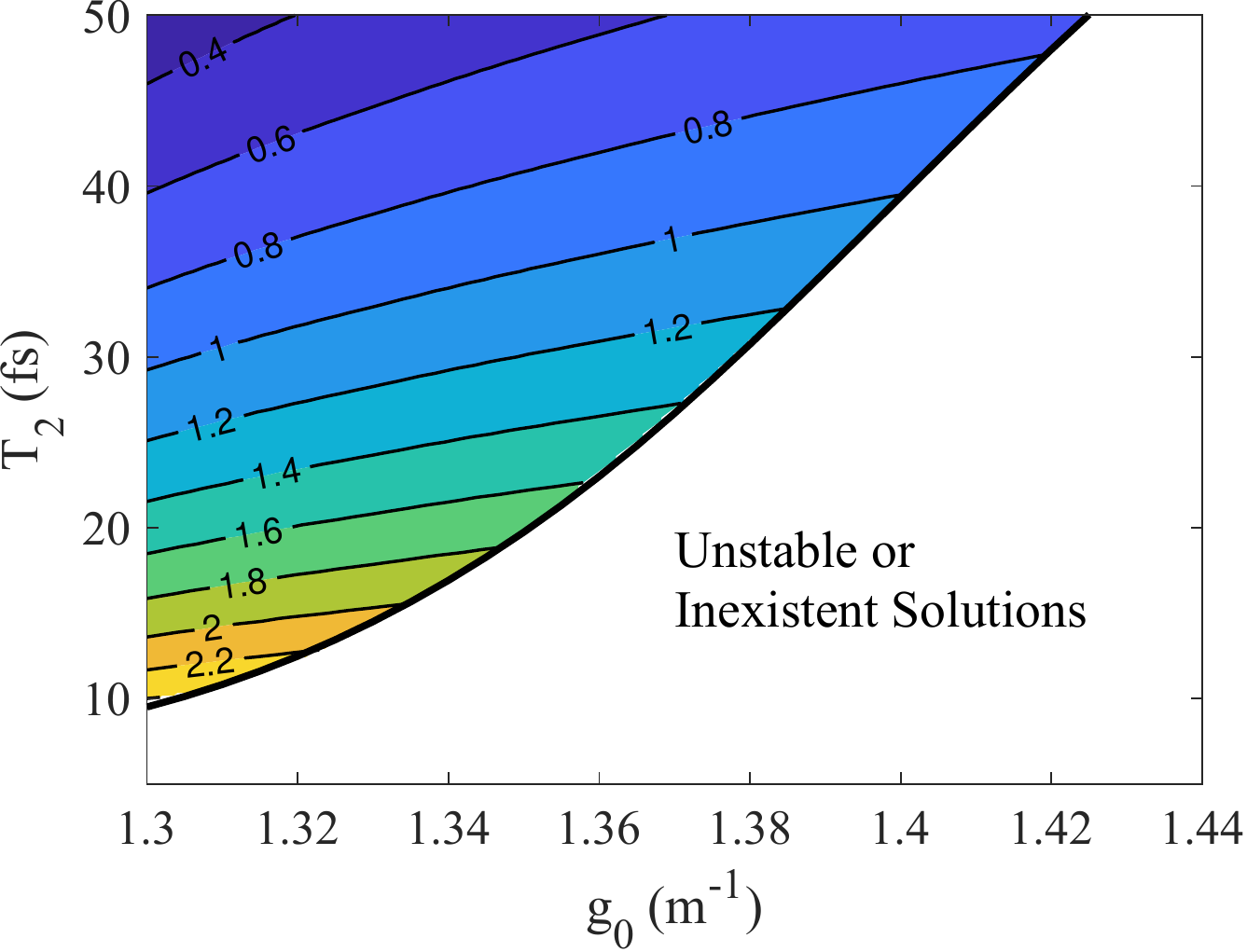}\label{fig:80stability_neg}}
    \subfigure[]{
    \centering\includegraphics[width = 0.48\linewidth]{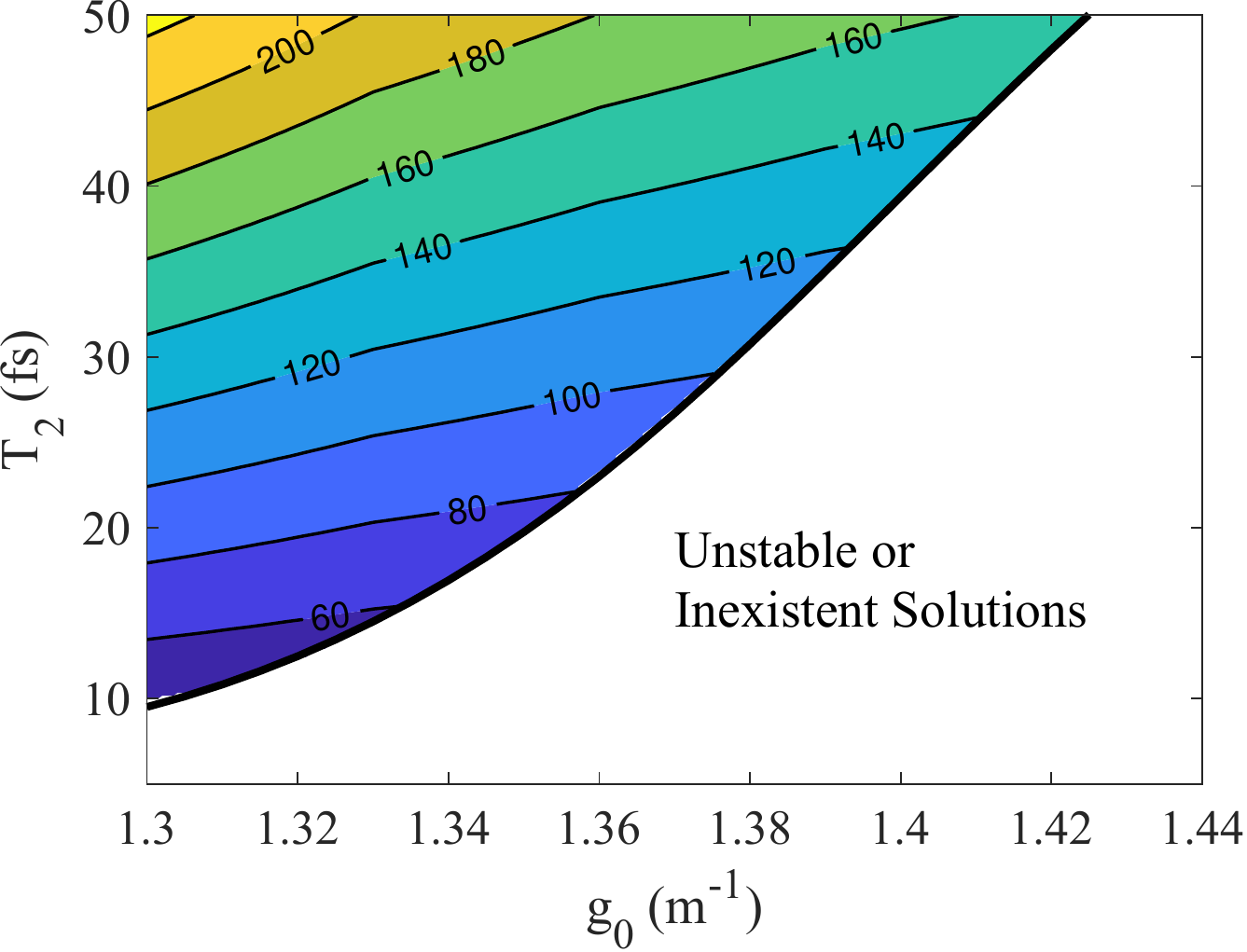}\label{fig:80stability_width}}

    \centering
    \subfigure[]{
    \centering\includegraphics[width = 0.48\linewidth]{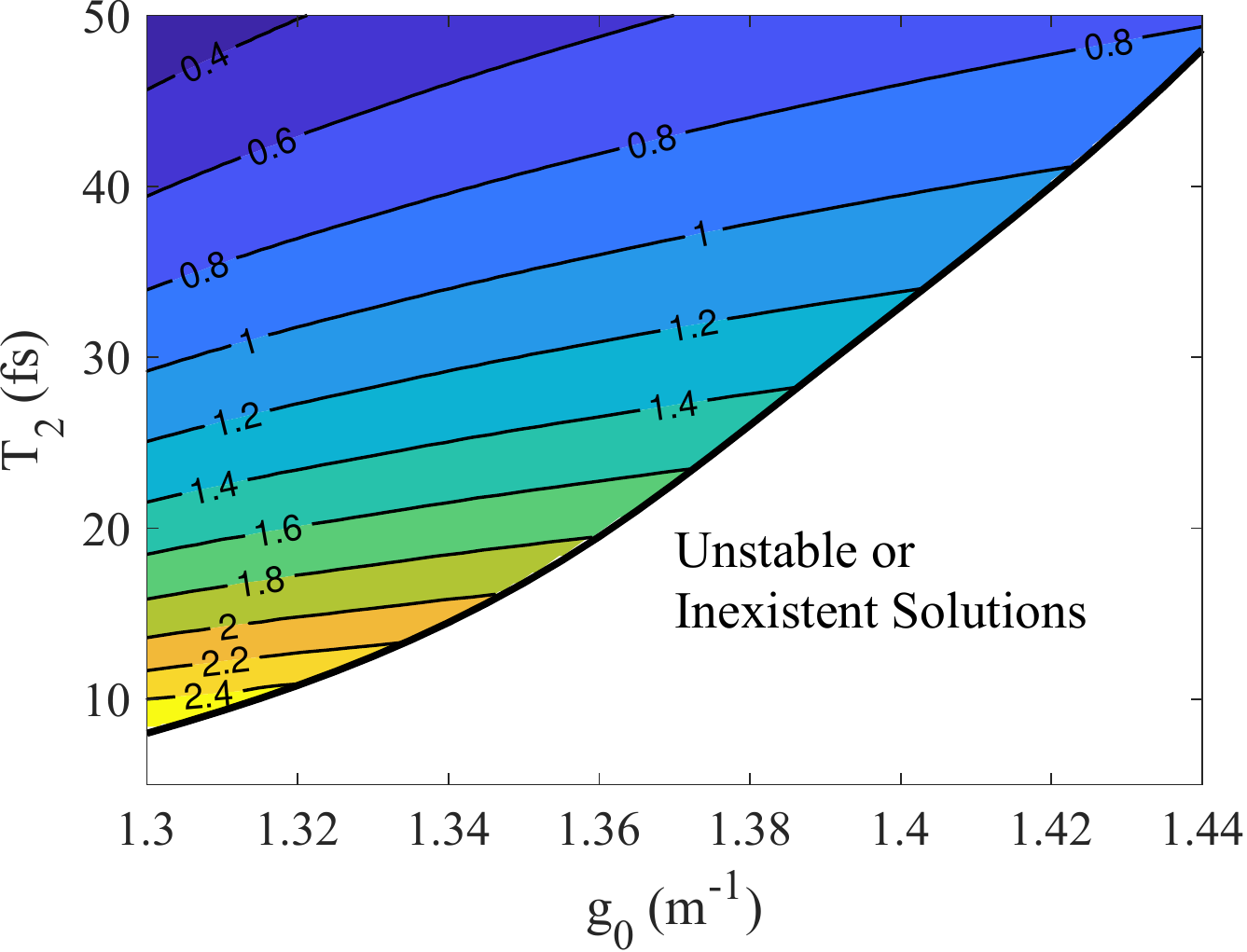}\label{fig:stability_neg}}
    \subfigure[]{
    \centering\includegraphics[width = 0.48\linewidth]{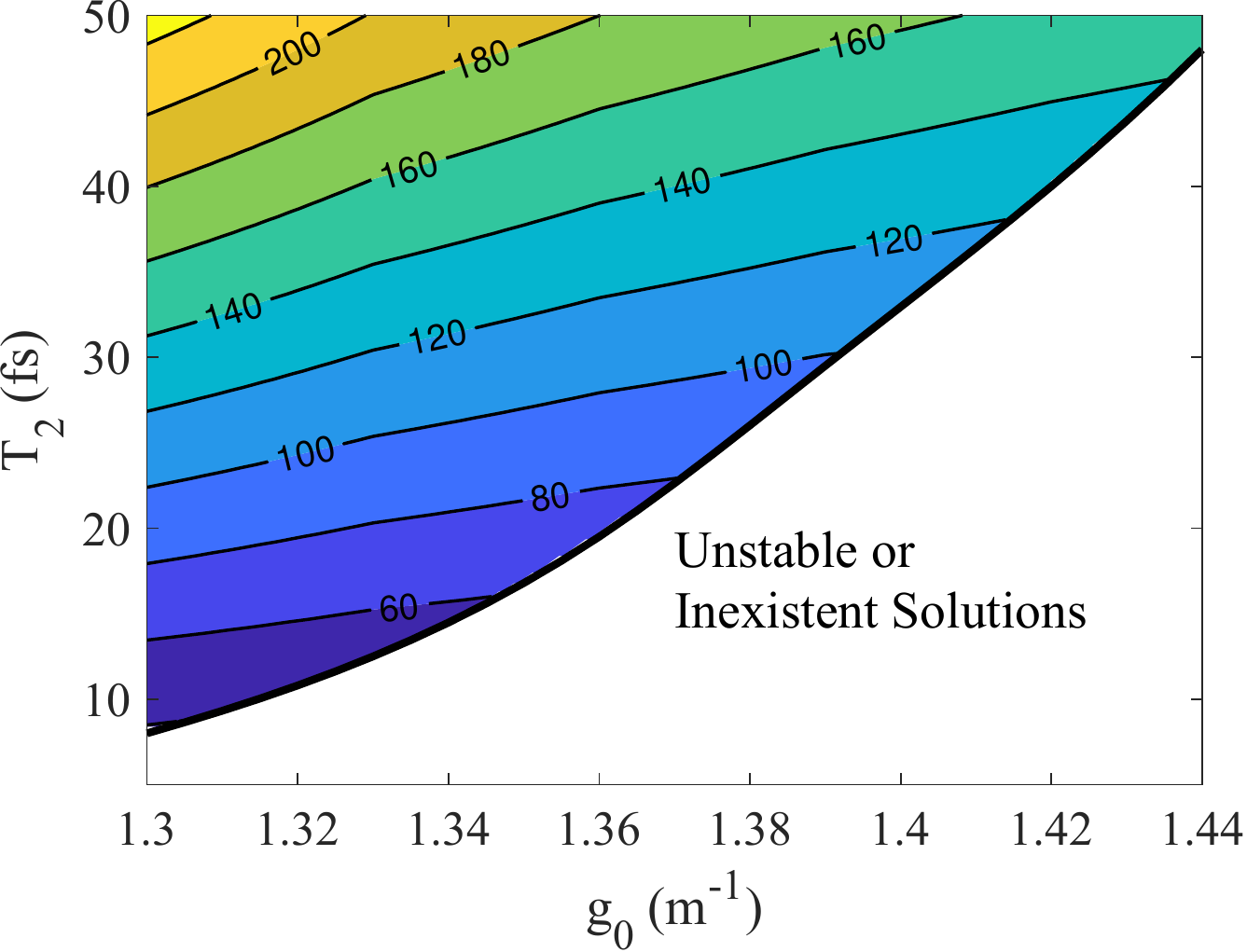}\label{fig:stability_width}}
   
    \caption{Stability border solitons in the $g_0-T_2$ plane. Above the curve, \subref{fig:80stability_neg} and \subref{fig:stability_neg} the base 10 logarithm of the energy in nJ, in \subref{fig:80stability_width} and \subref{fig:stability_width} the pulse width in fs are represented in a contourplot. Equation parameters: $\beta_2 = 0$, $\beta_4 = -0.08~\text{ps}^4\text{m}^{-1}$ and $P_\text{sat} = 80~\text{W}$ (top row) and $P_\text{sat} = 160~\text{W}$ (bottom row).}
    \label{fig:stability}
\end{figure}

In order to have a better understanding of the combinations of these parameters that lead to highly energetic pulses with short widths,
we searched for solutions and calculated stability eigenvalues on a $g_0-T_2$ region for $\beta_2 = 0~\text{ps}^2\text{m}^{-1}$ and $\beta_4 = -0.08~\text{ps}^4\text{m}^{-1}$. Fig. \ref{fig:stability} shows the energy (in logarithmic scale) and width (fs) values as contour plots for the stable solutions when $P_\text{sat}$ is $80~\text{W}$  and $160~\text{W}$.  
Note that, despite the energy increasing with the increase of $g_0$ for fixed $T_2$ and $P_\text{sat}$, which is in agreement with Fig. \ref{fig:E_G0_POS_FOD}, lower values of $g_0$ allow solutions to exist and evolve steadily for lower values of $T_2$, and solutions in that region have the highest energy.  Similarly, even though for fixed $T_2$ and $g_0$, a decrease of $P_\text{sat}$ was favourable for higher energy, if $T_2$ and $g_0$ are allowed to vary we may achieve higher energies for higher values of $P_\text{sat}$.
When it comes to the widths, represented in Fig. \ref{fig:stability_width}, it is shown that the pulse width decreases as $T_2$ decreases. $T_2$ being the parameter of the spectral filtering is influencing the pulse shape and, as expected, the pulse width is always changing in the same direction as $T_2$. Therefore, the most energetic pulses are also the shortest ones. In fact, using $T_2 = 7.5$ fs, $g_0 = 1.3~\text{m}^{-1}$ and $\beta_4 = -0.08~\text{ps}^4\text{m}^{-1}$ yielded quartic solitons with energy of 392 nJ and 39 fs width. These energy values are much greater than the ones presented in previous works \cite{tam19, Taheri2019, runge20}.

In conclusion, we found quartic soliton solutions of a mode-locked laser distributed model in the presence of positive and negative and 4OD, as well as in the absence or in the presence of negative GVD. The selection of laser parameters is of paramount importance for the existence and stability of solitons, as well for defining their shape. Most of profiles for negative 4OD have oscillations in their tails and for positive 4OD are hyperbolic secant in the center but upon a larger pedestal. Nevertheless, there exist profiles very similar to sech for both negative and positive 4OD if larger spectral filtering $T_2$ is considered. We studied the energy-width relation of the different solutions, finding that the energy dependence with the width varies with laser parameters. With positive 4OD, our results did not show a single universal trend, but with negative 4OD, most of our results followed an $\text{energy} \propto\text{width}^{-3}$ trend. The exceptions were for the sweep in $\beta_4$ as well as the sweep in $T_2$ for low $P_\text{sat}$ , where a non-monotonous trend was found. The highest energies and shortest widths were found for negative 4OD and for low $T_2$ values. To obtain solutions for such low $T_2$, $P_\text{sat}$ should be increased and $g_0$ decreased. In effect, our simulations of the distributed model show that the solution for surpassing current energy limits may be through quartic solitons by carefully optimizing gain bandwidth and spectral filtering within the available gain and saturation power levels.

\begin{backmatter}
\bmsection{Funding} Content in the funding section will be generated entirely from details submitted to Prism. Authors may add placeholder text in the manuscript to assess length, but any text added to this section in the manuscript will be replaced during production.

\bmsection{Disclosures} The authors declare no conflicts of interest.

\bmsection{Data availability} Data underlying the results presented in this paper are not publicly available at this time but may be obtained from the authors upon reasonable request.

\end{backmatter}

\bibliography{solitoes}

\bibliographyfullrefs{solitoes}

\end{document}